\documentclass{optica-article}

\journal{opticajournal} 

\articletype{Research Article}

\begin{document}

\title{Quantum noise in ranging with optical pulses}

\author{Mylenne Manrique,\authormark{1} Ilaria Gianani,\authormark{1} Marco Barbieri\authormark{1,3,4}, 
Valentina Parigi\authormark{2}, and Nicolas Treps\authormark{2}}

\address{\authormark{1}Dipartimento di Scienze, Universit\'a degli Studi Roma Tre, Via della Vasca Navale 84, 00146 Rome, Italy\\
\authormark{2}Laboratoire Kastler Brossel, Sorbonne Universit\'e, ENS-Universit\'e PSL, CNRS, Coll\`ege de France, 4 place Jussieu, 75252 Paris, France\\
\authormark{3} Istituto Nazionale di Ottica - CNR, Largo E. Fermi 6, 50125 Florence, Italy\\
\authormark{4} INFN, Sezione di Roma Tre, Via Della Vasca Navale 84, 00146 Roma, Italy}





\begin{abstract} 
Optical frequency combs combine ultrashort pulse duration and phase stability, making them powerful resources for high-precision ranging even when affected by atmospheric dispersion. It has been established that by classical modal engineering and mdoe-sensitive detection sensitivity to distance at the standard limit can be achieved, however attaining improved uncertainties by the use of squeezing has not been explored. Here, we apply an effective Hamiltonian framework to the problem of ranging with quantum frequency combs in order to derive the associated precision bounds for distance estimation. We analyse the role of intensity anti-squeezing and temporal beam shaping, and find that quantum solutions may be appealing mostly for short-distance applications.
\end{abstract}



Photonics is an invaluable platform for measuring distances in ranging and positioning problems~\cite{Liang2021,Peng2020,Cheng2019,Feng2022,Tan2022}. The flexibility provided by controlling different degrees of freedom of light yields a wide choice of techniques, each suitable for specific scenarios. Interferometric techniques are sought for their good accuracy, and the implementation of multi-wavelength schemes makes it possible to resolve ambiguities in determining absolute distances~\cite{li2025high,wu2018synthetic}. Time-of-flight measurements, instead, may warrant resolutions down to the nanometer scale~\cite{na2020ultrafast,Zhang2025}. Frequency combs bring these two aspects together and open the possibility of taking the best from both short pulse duration and phase stability~\cite{van2015mode,Wang2020,Lin2024,Chang2024,Camenzind2025}.

Light detection and ranging (LIDAR) works at its best when large distances are explored; problems emerge in geodesy\cite{glennie2013geodetic,An2024} and communication with satellites~\cite{younus2024overview}. Dispersion in the atmosphere, however, impacts the quality of the pulses, and it is almost always the case that monitoring the determining atmospheric conditions across the whole path remains beyond the experimenter's reach. The use of frequency combs offers a remarkable solution for this problem as well. In~\cite{jian2012real}, the performance of a homodyne detector, used as a phase-sensitive receiver, has been assessed. By inspecting the modal structure of the LiDAR pulses and that of the local oscillator, the authors have shown that pulse shaping can achieve sensitivity to a specific parameter, most crucially distance, while keeping the measurement insensitive to other quantities, such as humidity, temperature, and pressure~\cite{jian2012real}.   

The ongoing efforts in the development of quantum technologies have spurred interest in understanding fundamental limits in LiDAR applications in order to establish whether a quantum advantage may be achieved, with the yet elusive purpose of covert operation~\cite{Cohen2019,tham2024quantum,Ortolano2025}. However, the treatment of~\cite{jian2012real} is rooted in a classical description of light, as well as classical estimation theory for what concerns the retrieval of parameters. Moving to the quantum domain complicates the problem in both aspects, as it requires the quantisation of the field, and the adoption of the quantum estimation framework. Very recently, Gessner et al.~\cite{gessner2023estimation} have demonstrated how to tackle these issues by means of an effective Hamiltonian description associated with the modal structure of light. This has contributed a transformative approach to such problems, making them treatable and granting insight thanks to its analytical forms. This has since been generalised to the multiparameter case in~\cite{boeschoten2025estimation}. 

In this article, we apply the methods from Refs.~\cite{gessner2023estimation,boeschoten2025estimation} to obtaining fundamental limits in ranging with quantum frequency combs. We present a generalised treatment with respect to the one presented in~\cite{gessner2023estimation} for Gaussian pulse shapes. This is then applied to illustrate how intensity squeezing may modify the achievable precision. We find that the advantage is moderate, even in the ideal case. The origin of this effect can be traced to the dual aspect of the measurement - incorporating elements from the time-of-flight as well as from the phase-sensitive approaches. Our finding establish a benchmark for the use of quantum resources in LiDAR.

We consider a single-mode treatment in which a frequency comb, initially associated to a single spectral amplitude $A_0(\omega)$, propagates over a distance $L$. It is modified as $A(\omega)=A_0(\omega)e^{i \phi(\omega)}$, where $\phi(\omega)$ is the accumulated spectral phase:
\begin{equation}
\label{eq:dispersion}
    \phi(\omega) =\omega_0 t_\phi +(\omega-\omega_0)t_g+\frac{(\omega-\omega_0)^2}{\omega_0}t_{GVD},
\end{equation}
using a development up to the second order in $\nu=\omega-\omega_0$, the detuning with respect to a reference frequency $\omega_0$. The value of $L$ is then inferred indirectly from the measured dispersion, since its value is linked to the three parameters in~\eqref{eq:dispersion}, as $t_\phi = n_\phi L/c$,  $t_g =(n_\phi+\omega_0n_\phi')L/c$, and $t_{GVD}=\omega_0\left(n_\phi'+\omega^2n_\phi''/2\right)L/c$ . In these expressions $n_\phi$ is the refractive index of air at $\omega_0$, and the derivatives are taken with respect to frequency, thus invoking the group velocity and its dispersion. The optical frequency, however, is not the only factor determining the refractive index, as this is also dictated by ambient conditions. These need to be explicitly included in the treatment, by including the dependence of $n_\phi$ on them.  

In the language of metrology, the measurement procedure is therefore understood as a multiparameter estimation problem: one such parameter, the range $L$, is the relevant quantity of interest; the rest serve the purpose of describing other quantities determining $n_\phi$, thus adopting the approach of casting those unwanted effects as nuisance parameters~\cite{suzuki2020quantum,suzuki2020nuisance}.  The minimal variance is thus established by the quantum Cram\'er-Rao bound (QCRB) in matrix form. For the set of parameters $\{p_1, p_2,\dotsc\}$, the variance $\sigma^2_i$ of any unbiased estimator for $p_i$ is limited from below as $\sigma^2_i\geq (F^{-1})_{i,i}$, where $F$ is the quantum Fisher information matrix. This lower limit accounts for possible correlations existing among the parameters or, in other words, for the impact of nuisance on the estimation. 

Considering now the detection stage, estimation theory provides us with the result that homodyne detection performed with a local oscillator shaped with amplitude $A_{LO}(\omega)=\partial_{p_i} A_0(\omega)$ can achieve optimal sensitivity to the parameter $p_i$ when the others are known~\cite{jian2012real}. A second frequency comb is thus required as a local oscillator, with the possibility of acting on its temporal/frequency shape. Under these circumstances, the uncertainty is limited to a more optimistic bound: $\sigma^2_i\geq 1/F_{i,i}$. In the more realistic case in which the other parameter are actually unknown, different strategies should be put in place. The standard solution is that of considering a more complex measurement yielding values for all parameters. Alternatively,  the technique introduced in~\cite{jian2012real} prescribes a method for obtaining a different detection mode $A^{p_i}_{LO}$, which is insensitive to $p_j$ with $j\neq i$. Either solution leads to a sub-optimal measurement for $p_i$, and, remarkably, the associated loss of precision is always captured by the QCRB. 

The calculation of the quantum Fisher information matrix associated to the native parameters $t_\phi$, $t_g$  and $t_{GVD}$ is conveniently performed following the modal methods proposed in~\cite{gessner2023estimation}, and applied in~\cite{boeschoten2025estimation} to spectral measurements for the specific example of Gaussian pulses. A scalar product between two modes $f(\nu)$ and $g(\nu)$ is introduced as $(f,g)=\int f(\nu)^*g(\nu)d\nu$, and is used to introduce effective generator associated to each parameter as $H_k=i(A,A_k)a^\dag a$,  where $A_k=\partial_k A$ is the derivative with respect to the $k$-th parameter and $a$ is the annihilation operator associated to spectral mode $A$. The quantum Fisher information matrix $F$ is then obtained in terms of these generators; in particular, for pure states the relation is simply $F_{j,k}=4 \text{Cov}(H_j,H_k)$.  Further simplification can be carried out yielding to
\begin{equation}\label{eq:gessner}
    F_{i,j}=4 (A_i,A)(A,A_j)F_Q+4\text{Re}[(A_i,A_j)-(A_i,A)(A,A_j)]N,
\end{equation}
where $N$ is the average photon number in the initial mode, and
\begin{equation}
\label{eq:fq}
    F_Q = 4N^2-8\sum_{a,b\neq a} \left(\frac{1}{p_a}+\frac{1}{p_a}\right)^{-1}|\langle a|N| b\rangle|^2,
\end{equation}
for a quantum state $\rho = \sum_a p_a|a\rangle\langle a|$. This formalism is thus able to use the classical modal structure in order to describe the modification to the quantum state.  For pure states, the quantity $F_Q$ reduces to $F_Q=4\Delta^2N$, four times the photon number variance, hence, in this limit, the quantum Fisher information matrix is given by
\par\small
\begin{equation}
\label{eq:fim}
\begin{aligned}
    &F=\\
   &4\begin{pmatrix}
    \omega_0\Delta^2N& \omega_0 \mu_1 \Delta^2N& \mu_2\Delta^2N\\
    \omega_0 \mu_1 \Delta^2N & \mu_2N+\mu_1^2(\Delta^2N-N)& \frac{\mu_3}{\omega_0}N +\frac{\mu_1\mu_2}{\omega_0}(\Delta^2N-N)\\ 
   \mu_2\Delta^2N& \frac{\mu_3}{\omega_0}N +\frac{\mu_1\mu_2}{\omega_0}(\Delta^2N-N)& \frac{\mu_4}{\omega_0^2}N+\left( \frac{\mu_2}{\omega_0}\right)^2(\Delta^2N-N)
    \end{pmatrix}
    \end{aligned}
\end{equation} 
\normalsize
The coefficients $\mu_k$ denote the $k$-th order moments of the frequency distribution $\mu_k=\int(\omega-\omega_0)^k\vert  A(\omega-\omega_0)\vert^2d\omega$. The expression \eqref{eq:fim} thus enables to assess the influence of the mode shape, as well as of quantum fluctuations at once. Since this Fisher information matrix assumes a pure state, the increased photon number fluctuations occur as a consequence of a reduction of the uncertainty on the phase $\Delta^2\phi=1/\Delta^2N$. 

The QCRB based on \eqref{eq:fim}  is however unable to provide a limit to the precision on $L$ directly. A reparametrisation is thus needed in such a way to obtain a new Fisher information matrix $\tilde F$ in which $L$ appears in an explicit manner. Along with the distance, two more quantities are thus required, as the overall number of parameters must be taken. Following~\cite{jian2012real}, we consider the Edl\'en model for dispersion in air~\cite{edlen1966refractive} refined in~\cite{bonsch1998measurement}. The phase index $n_\phi$ is a function of the partial pressure of water $P_w$ and of a parameter $X$ encompassing the temperature $T$, the pressure $P$ and the concentration $x$ of $CO_2$:
\begin{equation}
\begin{aligned}
   n_\phi(\tilde \nu \vert X,P_w) = &1+10^{-8}\left( 8091.37+\frac{2333983}{130-\tilde \nu^2}+\frac{15518}{38.9-\tilde\nu^2}\right)X\\&-10^{-10}(3.802-0.0384\tilde \nu^2)P_w(\text{Pa}),
\end{aligned}
\end{equation}
where $\tilde \nu=1/\lambda$ is expressed in $\mu$m$^{-1}$. The parameter $X$ is defined as
\begin{equation}
\begin{aligned}
    \label{eq:parameterX}
    & X=\frac{P(\text{Pa})}{93214.60}\frac{1+10^{-8}(0.5953-0.009876 T(^\circ \text{C}))P(\text{Pa})}{1+0.0036610T(^\circ \text{C}) }\times\\
    &(1+0.5327(x-0.0004))
\end{aligned}
\end{equation}
The sought reparametrisation is thus the transformation $\{t_i\}=\{t_\phi,t_g,t_{GVD}\}\rightarrow \{c_j\}=\{L,X,P_w\}$. The new quantum Fisher information matrix  $\tilde F$  is obtained from $F$ by using the transposed Jacobi matrix with elements $B_{i,j}=\partial_{c_i}t_j$ : $\tilde F=BFB^{T}$~\cite{paris2009quantum}. The diagonal element of $\left(\tilde F^{-1}\right)_{jj}$  thus establish the relevant QCRB when considering the parameter set $c_j$. Notice that $B$ also accounts for correlation among the parameters $\{c_j\}$, which may not necessarily mirror those among $\{t_j\}$.

\begin{figure}
    \centering
    \includegraphics[width=\columnwidth]{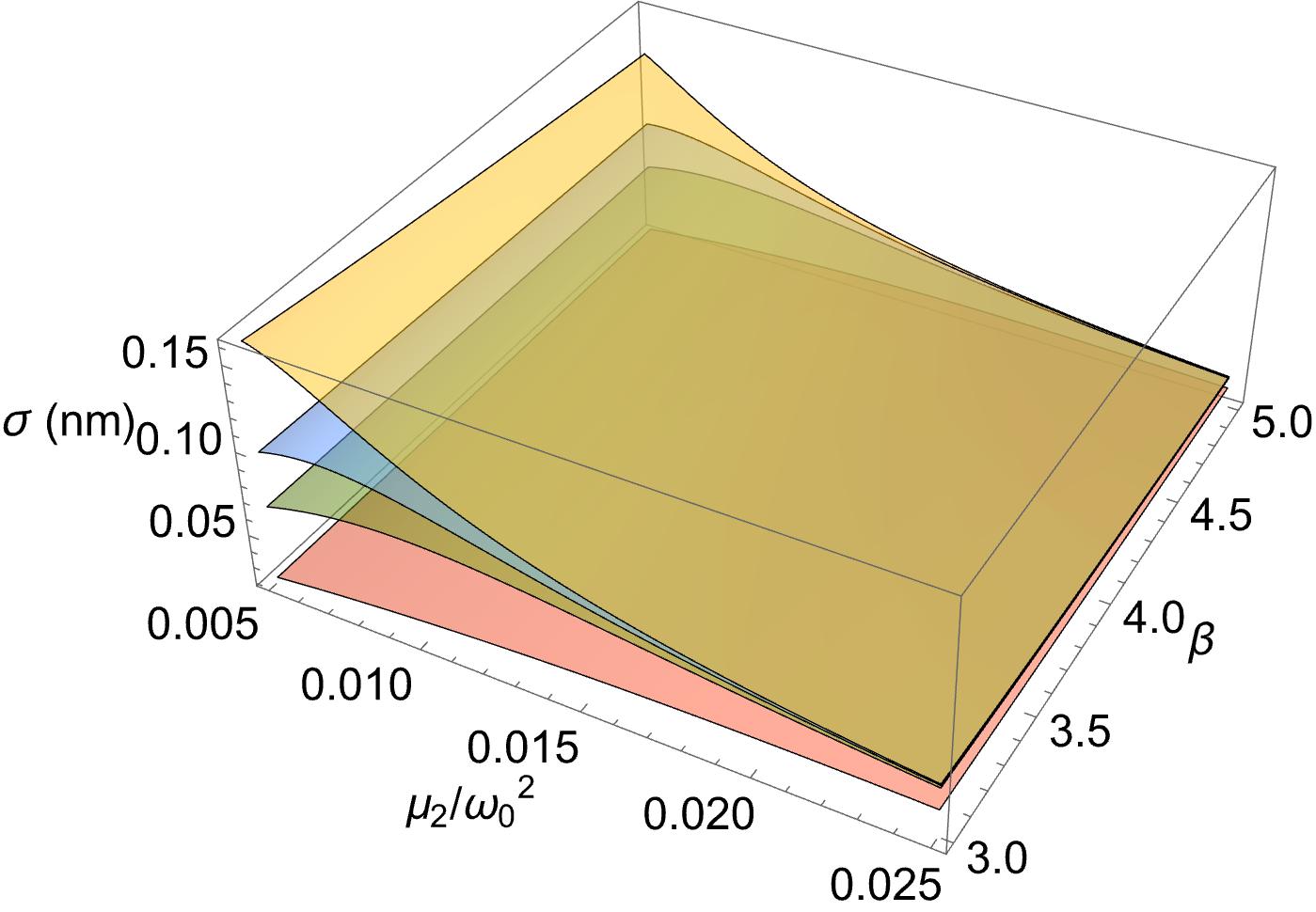}
    \caption{Uncertainty $\sigma$ of the retrieved value of $L$, using an average of $N=10^{16}$ photons in a symetric single mode, as a function of the second moment $\mu_2$ and kurtosys $\beta$ of the spectral distribution. The parameters are $\lambda_0=2\pi c/\omega_0=0.785$ nm,  $L=1$ km, $T=24^\circ$C, $P=1$ atm, $x=0.04\%$ , $P_w=0.0313$ atm. The four surfaces are associated to the shot noise, 3 dB intensity noise increase, 5 dB intensity noise increase, and 10 dB intensity noise increase, from above.}
    \label{fig:plot}
\end{figure}

We first consider symmetric pulses, thus odd moments are vanishing: $\mu_1 =0$, and $\mu_3=0$. The fourth-order moment is conveniently written in terms of the kurtosis $\beta=\mu_4/\mu_2^2$. Gaussian pulses, for instance, are characterised by $\beta=3$.  In Fig.~\ref{fig:plot} we report the value of the uncertainty $\sigma$ on $L$ as a function of the parameters $\mu_2$ and $\beta$ for a displaced squeezed vacuum $\hat D(\alpha)\hat S(re^{i\varphi})\vert 0\rangle$: its average photon number is $N=\vert \alpha\vert^2+\sinh^2(r)$, and the variance is $\Delta^2N=e^{2r}\vert\alpha\vert^2+\sinh(2r)/2$, when optimising over the phase $\varphi$, viz. $\varphi=\pi+2\arg(\alpha)$. 
 
The results indicate that intensity anti-squeezing is most effective when relatively long pulses are used, and that, in this same regime, there is an advantage in adopting spectra with larger kurtosis, such as the usual hyperbolic secant expression for soliton pulses. Overall, the most effective strategy is to shorten the pulse in time, signalling that the aspects linked to the time-of-flight measurement are prominent in determining the precision. We can gain further intuition for such a behaviour by inspecting the variances at the quantum CRB set by \eqref{eq:fim}
\begin{align}
&\Delta^2t_\phi=\frac{1}{4\omega_0^2}\left(\frac{\mu_2^2}{\mu_4-\mu_2^2}\frac{1}{N}+\frac{1}{\Delta^2N}\right)=\frac{1}{4\omega_0^2}\left(\frac{1}{\beta-1 }\frac{1}{N}+\frac{1}{\Delta^2N}\right),\\
    &\Delta^2t_g=\frac{1}{4\mu_2 N},\\
    &\Delta^2t_{GVD}=\frac{\omega_0^2}{4(\mu_4-\mu_2^2)}\frac{1}{N}=\frac{\omega_0^2}{4\mu_2^2(\beta-1)}\frac{1}{N}.
\end{align}
The photon number variance only determines the uncertainty on $t_\phi$, while the others are governed by the shot noise level, {i.e.} by the mean photon number. Acting at the quantum level by controlling noise is bound to show limited effects.

\begin{figure}
    \centering
    \includegraphics[width=\columnwidth]{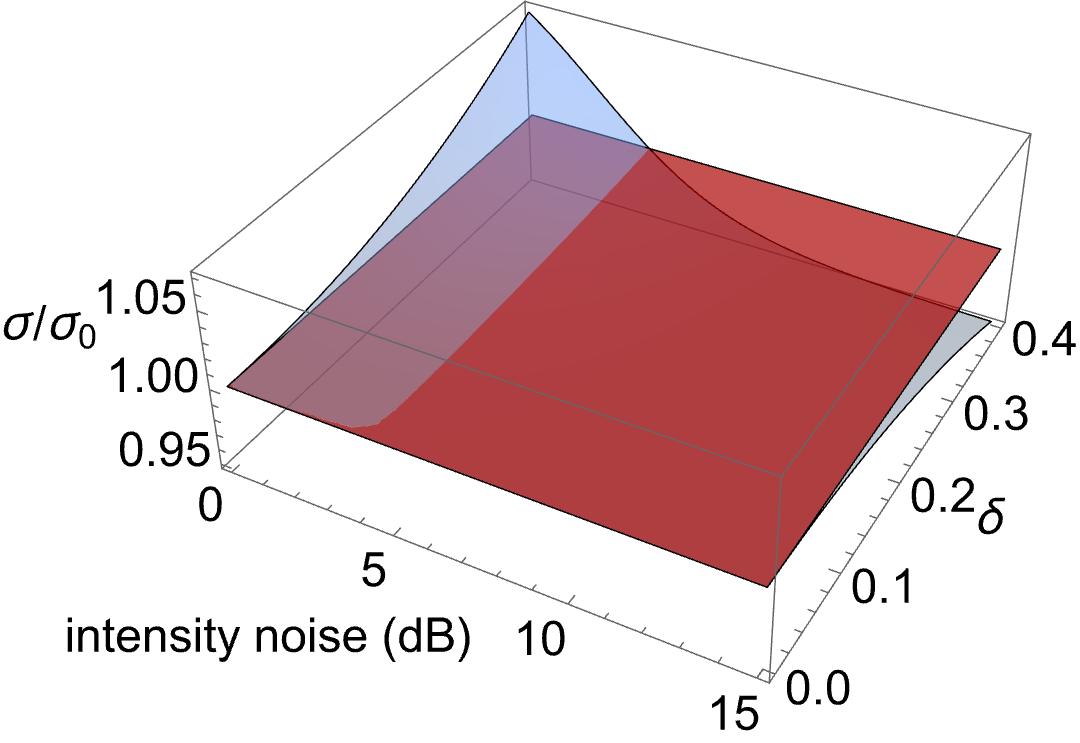}
     \caption{Uncertainty $\sigma$ on the retrieved value of $L$ in as a function of the intensity squeezing and of the asymmetry parameter $\delta$ of a skewed distribution, normalised to that from a Gaussian pulse. We chose $\mu_2=(\omega_0/10)^2$, all other parameters are the same as in Fig.~\ref{fig:plot}. }
    \label{fig:asymmetry}
\end{figure}

The asymmetry of the pulse spectrum can also be accounted for using the same formalism. In Fig.\ref{fig:asymmetry} we show the uncertainty $\sigma$ on $L$  (corresponding to $r=0$) for pulses following a skewed-normal distribution with an asymmetry parameter $\delta$; in general, the parameter $\delta$ ranges in $-1< \delta < 1$, with $\delta=0$ corresponding to the standard Gaussian distribution, and $\pm 1$ are asymptotic values for infinitely skewed distributions. When comparing this uncertainty to that from a Gaussian pulse $\sigma_0$, we can observe that the impact of asymmetry remains limited. Notably, with moderate squeezing the performance is slightly worsened, whereas it improves at higher level. This is another signature of the interplay between quantum and modal aspects in this class of estimation problems.

\begin{figure}
    \centering
    \includegraphics[width=\columnwidth]{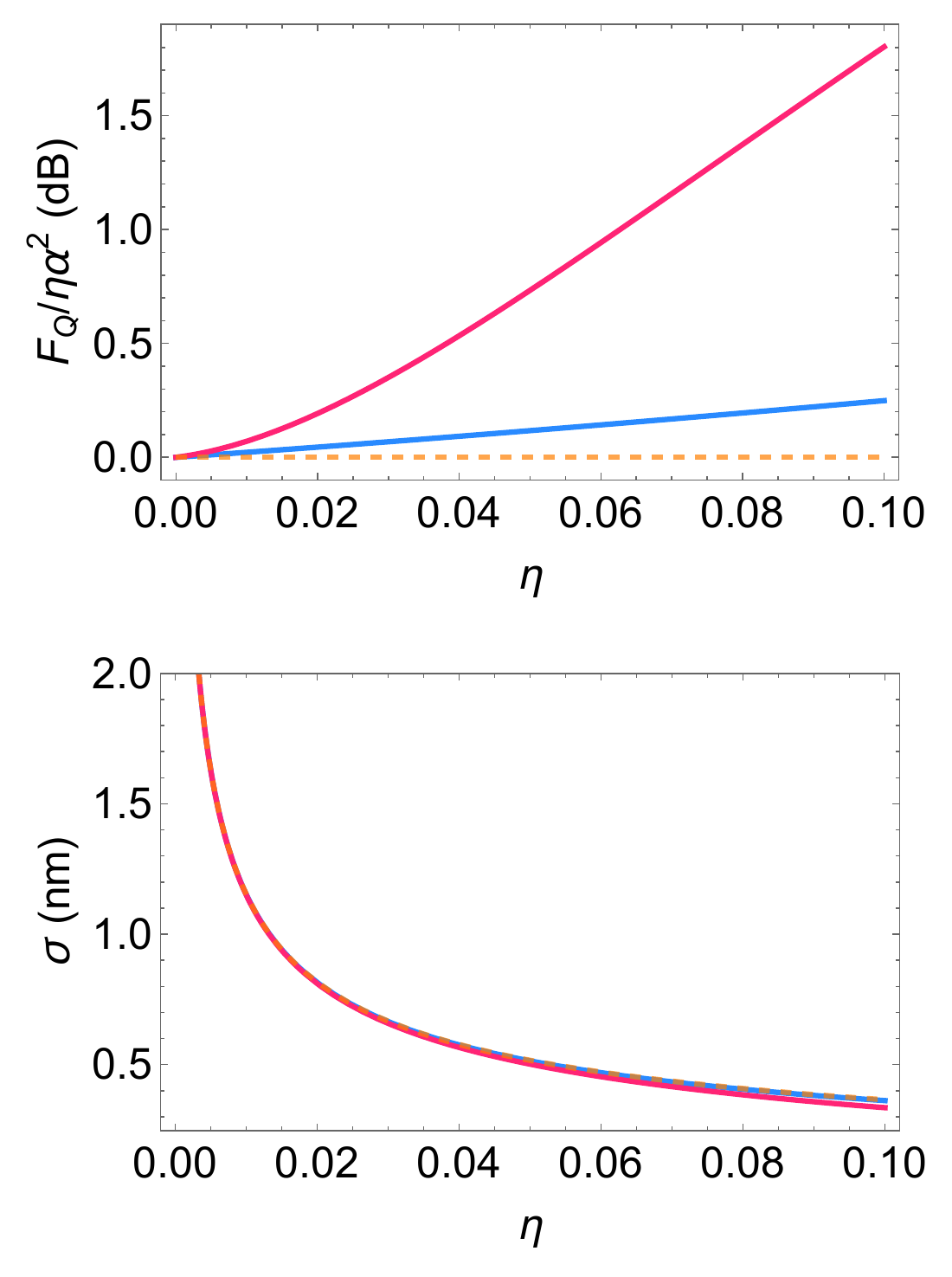}
     \caption{Effect of loss on the estimation. Upper panel: $F_Q$ for the state affected by a channel with transmittance $\eta$.
     Lower panel: uncertainity $\sigma$ on $L$. In both panels, the yellow dashed curve corresponds to the shot noise with an initial mean photon number $N=10^{16}$, blue solid curve 3dB of initial intensity noise squeezing, red solid curve with 10dB of initial intensity noise squeezing. We chose a Gaussian pulse with $\mu_2=(\omega_0/10)^2$, all other parameters are the same as in Fig.~\ref{fig:plot}. }
    \label{fig:loss}
\end{figure}

The use over long distances normally imposes high loss in the system; in the intended detection scheme, these are not only determined by optical attenuation, but also by the modal matching with the local oscillator. The formulae above thus need to be changed in order to account for the mixing in the state, hence $F_Q$ requires its full expression~\eqref{eq:fq}. The results are shown in Fig.~\ref{fig:loss}, indicating that, at low transmission, the use of intensity squeezing becomes practically irrelevant, although a change of $F_Q$ with respect to the shot noise can still be appreciated.

In this work, we have investigate a general framework for assessing the fundamental limits of quantum‑enhanced ranging with frequency combs. Our results deliver a realistic, however sober, appraisal of their potential and, especially, their limitations. Since modal parameters dictate the problem, it manifests as a remarkable mixture of classical and quantum aspects in the estimation, thus lending itself to be tackled by the methods of~\cite{gessner2023estimation}. Our analysis emphasises how the practical impact of intensity squeezing is far more restrained than often anticipated. In detail, the dual nature of comb-based LiDAR combining pulsed structure and phase coherence limits how well squeezing can suppress inherent fluctuations of phase.
As a result, even under near-ideal conditions, the conceivable quantum advantage remains modest. These considerations bear even greater importance once loss and asymmetries in the modal structures are included. These imply that, in long-distance applications where these effects are unavoidably, the practical usefulness of squeezing vanishes.

This suggests applications in the complementary domain of short-range scenarios - these address more controlled environments, such as in 3D scanning of artworks, as well as open systems like the navigation of self-driving vehicles. In those cases loss can be kept at more manageable methods, and  atmospheric dispersion does not bear the same impact. The modal dependence we identified further suggests that the short pulse regime, which increases the relevance of the time‑of‑flight contribution to the information, provide  more favourable operations.

\section*{Appendix}
In the Edl\'en model, the reparametrisation matrix is given by 
\begin{equation}
\begin{aligned}
&B=\\
&\begin{pmatrix}
k_0
+k_1 X 
+k_2 P_w
&
k_0
+ k_3 X
+ k_4 P_w
&
k_5 X
+ k_6 P_w
\\
k_1 L
&
k_3 L
&
k_5 L
\\
k_2 L
&
k_4 L
&
k_6 L
\end{pmatrix}
\end{aligned}
\end{equation}
\normalsize
with $k_0=3.34$, $k_{1}=+ 8.90\times 10^{-4}$, $k_{2}=- 1.25\times 10^{-9}$, $k_3=9.07\times 10^{-4}$, $k_4=- 1.21\times 10^{-9}$, $k_5=2.53\times 10^{-4}$, and $k_6=6.24 \times 10^{-11}$, adopting the same units as described in the main text. 

The skew-normal distribution is defined, up to normalisation, as: $P(z;\alpha)=e^{-\frac{z^2}{2}}(1+\text{erf}(\alpha z/\sqrt{2}))$, where $z$ is the standard normal variable. The asymmetry parameters is $\delta =\alpha/\sqrt{1+\alpha^2}$. 

\section*{Acknowledgment}
Special thanks to Alvaro Bovetti and Camilla Tartoni for insightful discussions. This work is supported by by the PRIN 2022 MUR Project EQWALITY (N. 202224BTFZ) IG and MB acknowledge support from MUR Dipartimento di Eccellenza 2023-2027.

\bibliography{sample}

\end{document}